\documentstyle[twocolumn,aps,epsf]{revtex}
\renewcommand{\vec}[1]{\mbox{\boldmath$#1$}}
\newcommand{\mat}[1]{\mbox{\sf #1}}
\begin{document}
\twocolumn[\hsize\textwidth\columnwidth\hsize\csname
@twocolumnfalse\endcsname
\title{Conductance Quantization and  Magnetoresistance in Magnetic Point Contacts}
\author{Hiroshi Imamura} \address{CREST and Institute for Materials
Research, Tohoku University, Sendai 980-8577, Japan} \author{Nobuhiko
Kobayashi} \address{RIKEN (The Institute of Physical and Chemical
Research), Wako, Saitama 351-0198, Japan} \author{Saburo Takahashi and
Sadamichi Maekawa} \address{Institute for Materials Research, Tohoku
University, Sendai 980-8577, Japan} \maketitle
\begin{abstract}
  We theoretically study the electron transport through a magnetic
  point contact (PC) with special attention to the effect of an atomic
  scale domain wall (DW). The spin precession of a conduction
  electron is forbidden in  such an atomic scale DW and the sequence
  of quantized conductances depends on the relative orientation of
  magnetizations between left and right electrodes.
  The magnetoresistance is strongly enhanced for the narrow PC and
  oscillates with the conductance.
\end{abstract} 
\vskip1pc]

The electron transport through metallic nanostructures such as
nanowires and nanoparticles has attracted much attention.
The quantized conductances of integer
multiples of $e^{2}/h$ have been observed  
in metallic nanowires, where an atomic scale
constriction called point contact (PC) is made by pulling off two
electrodes in contact \cite{krans94,olesen94,rubio96,costa97}.
The conductance quantization is
well explained by Landauer's formula\cite{buttiker85}
of quantum ballistic transport combined with the adiabatic
principle\cite{glazman88,yacoby90}: for a nonmagnetic PC, the spin-up
and spin-down electrons make the same contribution and the unit of the
conductance quantization is $2e^{2}/h$.

However, if the PC is made of a magnetic-metal such as Fe, Co and Ni,
the exchange energy removes the spin degeneracy of conduction
electrons and an atomic scale domain wall (DW) is created in the 
PC \cite{bruno99}.
The conductance depends on the relative orientation of
magnetizations between left and right electrodes, like magnetic tunnel
junctions\cite{julliere75,maekawa82,hima_fnf}.  The spin dependent
transport such as the tunnel magnetoresistance (TMR) in magnetic
tunnel junctions and giant magnetoresistance (GMR) in magnetic
multilayers\cite{baibich88} is of current interest both in fundamental
physics and application in spin-electronics.  It is then intriguing to
ask how the exchange energy and the DW affect the
electron transport in a magnetic PC.  Recently, the following
fascinating experimental results have been reported 
in Ni PC: the magnetoresistance in excess of 200\%\cite{garcia99}, the
spin-dependent conductance quantization\cite{oshima98} and the
$2e^{2}/h$ to $e^{2}/h$ switching of the quantized
conductance\cite{ono99}.

\begin{figure}
   \epsfxsize=0.95\columnwidth \centerline{\hbox{ \epsffile{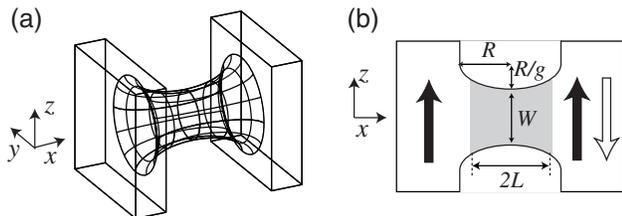}}
     }
    \caption{ (a) The geometry of the constriction potential
     $V(r_{\parallel},x)$ is schematically shown. (b) The crosssection
       of the constriction potential in $xz$ plane. Arrows represent the
       magnetization vectors of electrodes. In the right electrode,
       the filled (hollow) arrow represents the magnetization vector
       for the F (A) alignment.  The shaded region in the
       constriction represents the DW with thickness $2L$ for the A
       alignment.}
   \label{fig:pot}
 \end{figure}

In this Letter, we study the electron transport through a magnetic PC
shown in Figs. \ref{fig:pot} (a) and \ref{fig:pot} (b) by using the
recursion-transfer-matrix (RTM)
method\cite{hirose95,brandbyge97j,nkoba99}.
When a magnetic field is applied, the magnetizations of left and right
electrodes are parallel.  In this
ferromagnetic (F) alignment, the quantized conductance of odd integer
multiples of $e^{2}/h$ appears, since the spin-up and spin-down
electrons contribute to the conductance in the different way.
On the other hand, in the absence of the magnetic field, the system is
in the antiferromagnetic(A) alignment, where the magnetizations of
left and right electrodes are antiparallel. 
The DW is created inside the constriction.  We show that the spin
precession of conduction electrons is forbidden in
such an atomic scale DW.   The contribution
to the conductance from spin-up and spin-down electrons are the
same and the unit of the conductance quantization is $2e^{2}/h$.
We also show that the magnetoresistance (MR) is strongly enhanced for
the narrow PC and oscillates with the conductance.  Our
study explains the recent experimental results
\cite{garcia99,oshima98,ono99} and provides a new direction for
spin-electronic devices: an atomic scale DW.

We consider the system consisting of two flat electrodes connected by
a contact as shown in Figs. \ref{fig:pot} (a) and \ref{fig:pot} (b).
The effective one-electron Hamiltonian is given
by\cite{slonczewski89},
\begin{equation}
  {\cal H}=-\frac{\hbar^{2}}{2m}\nabla^{2} + V(r_{\parallel},x)
-\vec{h}(x)\cdot\vec{\sigma},
  \label{eq:hamiltonian1}  
\end{equation}
where $V(r_{\parallel},x)$ is the constriction potential, $\vec{h}(x)$
is the exchange field and $\vec{\sigma}$ is the Pauli spin matrix.
Here, $\vec{r}_{\parallel}=(y,z)$ is the position vector in the $yz$
plane.
We define the constriction potential $V(r_{\parallel},x)$ as
\cite{brandbyge97j},
\begin{equation}
  V(r_{\parallel},x)\!=\!V_{0}\left\{
    \begin{array}{l}
      \theta(R+x)\theta(R-x) \\ \vspace{1em} \mbox{\hspace{13ex}for
      $r_{\parallel} \ge R/g+ W/2$}\\ \theta(R-\sqrt{x^{2} + (R + gW/2
      - gr_{\parallel})^{2}})\\\mbox{\hspace{13ex}for $r_{\parallel}
      < R/g+ W/2$},
    \end{array}\right.
  \label{eq:constriction}
\end{equation}
where $V_{0}$ is the height of the potential, $R$ is the radius of the
elliptic envelope with a deformation parameter $g$ and $W$ is the
width of the constriction. 
We take the $z$ axis to be parallel to the
exchange field in the left electrode.  For the F
alignment, the exchange field is constant $\vec{h}(x)=(0,0,h_{0})$.  
For the A alignment, on the contrary, $\vec{h}(x)$ is not constant
inside the DW.  In the classical theory, $\vec{h}(x)$ is given by
\cite{bruno99,levy97}
\begin{eqnarray}
  \! h_{x}(x)&\!=\!&0,\ \     
  h_{y}(x)\!=\!h_{0} \theta(L-|x|)\cos(Q x ),\nonumber\\
  \! h_{z}(x)&\!=\!&h_{0}\{ \theta(\!-L\!-\!x)\!-\!\theta(\!-\!
    L\!+\!x) \!-\!\theta(L\!-|x|)\sin(Q x)\},
    \label{eq:hz}
\end{eqnarray}
where $L$ is half the thickness of the DW and $Q=\pi/2L$. 
As pointed out by Bruno\cite{bruno99}, the
DW is about the size of the PC. For such an atomic scale DW,
we have to derive the exchange field $\vec{h}(x)$ on the basis of the
quantum theory.  As we will show later, the $y$ component of the
exchange field vanishes $h_{y}(x) = 0$ and the spin of conduction
electrons cannot rotate in the atomic scale DW. 

In order to obtain the stationary scattering states, we employ the RTM
method\cite{hirose95} including the spin degree of freedom.  A
two-dimensional supercell structure is considered in the $y$ and $z$
directions.  We take a unit of the supercell large enough to regard
the transmission through the supercell as that in the non-periodic
potential.  Owing to the Bloch's theorem, electronic states are
written in terms of the discrete reciprocal lattice vectors
$\vec{K}^{n}_{\parallel}$ in the $y$ and $z$ directions.  The $m$th
stationary scattering state with spin $\sigma$ is written as,
\begin{equation}
  \psi_{m\sigma}(\vec{r}_{\parallel},x) =
  e^{i\vec{k}_{\parallel}\cdot\vec{r}_{\parallel}}
  \sum_{n\sigma^{\prime}} \phi_{n\sigma^{\prime},m\sigma}(x)
  e^{i\vec{K}_{\parallel}^{n}\cdot\vec{r}_{\parallel}},
\end{equation}
where $\vec{k}_{\parallel}$ is the conserved Bloch $k$-vector and
$\phi_{n\sigma^{\prime},m\sigma}(x)$ are unknown coefficients to be
solved.  The combination of the index for the reciprocal lattice
vector and the spin, $(n,\sigma)$, defines the ``channel''.  The
number of channels $N_{c}$ is truncated by including only the set of
$\vec{K}_{\parallel}$ satisfying $| \vec{k}_{\parallel} +
\vec{K}^{n}_{\parallel} |^{2} < (2mE_{\rm c})/ \hbar^{2}$, where
$E_{\rm c}$ is the cutoff energy.

Outside the left (right) boundary of the scattering region, electrons
with spin $\sigma=\uparrow,\downarrow$ feel the constant potential,
$U^{\downarrow}_{L}=U^{\uparrow}_{R}=h_{0}$,
$U^{\uparrow}_{L}=U^{\downarrow}_{R}=-h_{0}$ and
$\psi_{m\sigma}(\vec{r})$ is written as,
\begin{equation}
\!\!\!\!\psi_{m\!\sigma\!}(\!\vec{r}\!) \!\! =\!\! \left\{
    \begin{array}{ll}
      \!\!e^{i\theta_{\!m\sigma}}e^{ik_{\! m \sigma}^{L} x}
      e^{i(\vec{k}_{\!\parallel} +
      \vec{K}_{\!\parallel}^{\!m})\cdot\vec{r}_{\!\parallel}}&\\
      \vspace{1em} \!\!\!+ \! \! \sum_{\!n \sigma^{\prime}}
      \!\!r_{\!n\sigma^{\prime}\!\!,m\sigma}
      e^{\!-ik_{\!n\sigma^{\prime}}^{L} \!x}
      \!e^{i(\vec{k}_{\!\parallel} +
      \vec{K}_{\!\parallel}^{\!n})\cdot\vec{r}_{\!\parallel}} &
      \!\mbox{for} x \!\le\! x_{\!L} \\ \!\!\sum_{\!n\sigma^{\prime}}
      \!t_{\!n\sigma^{\prime}\!\!,m\sigma}
      e^{\!-ik_{\!n\sigma^{\prime} }^{R} \!x}
      \!e^{i(\vec{k}_{\!\parallel} +
      \vec{K}_{\!\parallel}^{\!n})\cdot\vec{r}_{\!\parallel}} &
      \!\mbox{for} x\!\ge\! x_{\!R}, \\
    \end{array}\right. 
  \label{eq:ini_cond}
\end{equation}
where $r_{n\sigma^{\prime}\!\!,m\sigma}$
($t_{n\sigma^{\prime}\!\!,m\sigma}$) is the reflection (transmission)
coefficient, $\theta_{m\sigma}$ is the initial phase,
$k_{n\sigma^{\prime}}^{L(R)} =
\{(2m/\hbar^{2})(E_{F}-U_{L(R)}^{\sigma^{\prime}}) -
|\vec{k}_{\parallel} + \vec{K}_{\parallel}^{n}|^{2}\}^{1/2}$ and
$x_{L(R)}$ is the left(right) boundary of the scattering region.  The
coefficients $r_{n\sigma^{\prime}\!\!,m\sigma}$ and
$t_{n\sigma^{\prime}\!\!,m\sigma}$ are obtained by solving the RTM
equation.  The transmission matrix of the system, $\mat{T}$, is
expressed in terms of the coefficients
$t_{n\sigma^{\prime}\!\!,m\sigma}$ as
$ \mat{T} = (\mat{k}_{R}^{\frac{1}{2}})^{T}\ \mat{t}\
\mat{k}_{L}^{-\frac{1}{2}}$,
where $\mat{k}_{R}^{\frac{1}{2}} (\mat{k}_{L}^{-\frac{1}{2}})$ is a
$N\times M$ rectangular matrix whose $i,j$ elements are given by
${k_{i}^{R}}^{\frac{1}{2}}\delta_{i,j}
({k_{i}^{L}}^{-\frac{1}{2}}\delta_{i,j})$ and $M$ is the number of
the open channels deep in the right (left) electrode.  The conductance
per supercell is calculated as
\begin{equation}
  G=\frac{e^{2}}{h}\int {\rm d}\vec{k}_{\parallel} \frac{{\cal
  S}}{(2\pi)^{2}}{\rm tr}\left(\mat{T}^{\dag}\mat{T}\right),
\label{eq:conductance}
\end{equation}
where ${\cal S}$ is the area of the supercell.
For sufficiently large $yz$ unit cell, it is sufficient to
use only $\Gamma$ point
$\vec{k}_{\parallel}=0$\cite{brandbyge97j,nkoba99}.

We choose the commonly accepted values of the material parameters
for typical ferromagnetic metals of Ni, Co and Fe\cite{levy97,gregg96}.
The Fermi energy and the height of the constriction potential are
taken to be $E_{F}=3.8$ eV $(k_{F}= 1.0 {\rm \AA}^{-1})$ 
and $V_{0}=2E_{F}$, respectively.  
The length of the constriction and the thickness of the DW are taken
to be $2R=2L=10$ \AA\ corresponding to $3\sim 5$ atoms and the
deformation parameter is $g=10$.
We choose the magnitude of the
exchange field $h_{0}=0.3 \sim 0.7$ eV and replace the step function
$\theta$ in Eq. (\ref{eq:constriction}) by the Fermi function with a
width of 0.25 \AA\  to make the constriction potential smooth.  
We use the square supercell with
linear dimension of $20$ \AA\ in the $y$ and $z$ directions.  The
left(right) boundary of scattering region is taken to be
$x_{L(R)}=-10$ \AA\ $(10 $\AA $)$ and the size of the mesh is 0.2
\AA\ in the $x$ direction.  The cutoff energy is $E_{\rm c} = 21.8$ eV
and 354 channels are used in the numerical calculation.

\begin{figure}
  \epsfxsize=0.95\columnwidth \centerline{\hbox{ \epsffile{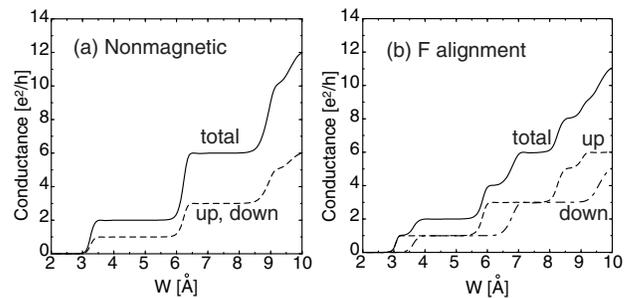}}
    }
  \caption{
    (a) The conductance curves for the nonmagnetic PC, $h_{0}=0$, are
    shown.  The total conductance is plotted by the solid line.  The
    conductances for spin-up and
    spin-down electrons are degenerate and plotted by the dotted line.
    (b) The conductance curves for the F alignment are shown.
    The total conductance is indicated by the solid line. The
    conductance for spin-up (spin-down) electrons is plotted by the
    dashed (dot-dashed) line. The exchange field is $h_{0}=0.5$ eV.
}
  \label{fig:PF}
\end{figure}

The conductance quantization of a nonmagnetic PC has been studied by
several
authors\cite{glazman88,yacoby90,brandbyge97j,nkoba99,lang95,brandbyge97d,cuevas98f}
and the sequence of quantized conductances, even integer multiples of
$e^{2}/h$, is well understood in the adiabatic
picture\cite{glazman88,yacoby90}.  In Fig.\ref{fig:PF}
(a), we show the conductance curve for a nonmagnetic PC.  The missing
of the plateau at 4 $e^{2}/h$ and 8 $e^{2}/h$ is due to the rotational
symmetry of the constriction potential\cite{brandbyge97j}. 
For a magnetic PC, the degeneracy of the spin-up and spin-down
conductances is removed by the exchange energy and plateaus of odd
integer multiples of $e^{2}/h$ appear for the F alignment as shown in
Fig. \ref{fig:PF} (b).  This kind of the spin-dependent conductance
quantization was first observed in semiconductor PCs under high
magnetic fields\cite{wharam88,wees88}, where the spin degeneracy is removed by
the Zeeman energy.  The key point is that the width of the
constriction $W$ at which the number of transmitting channels changes
is spin-dependent, because the Fermi wavelength is different between
spin-up and spin-down electrons.
The width $W$ at which the new transmitting
channel opens for spin-up (spin-down) electrons  decreases
(increases) as the exchange filed  increases.

\begin{figure}
  \epsfxsize=0.95\columnwidth \centerline{\hbox{ \epsffile{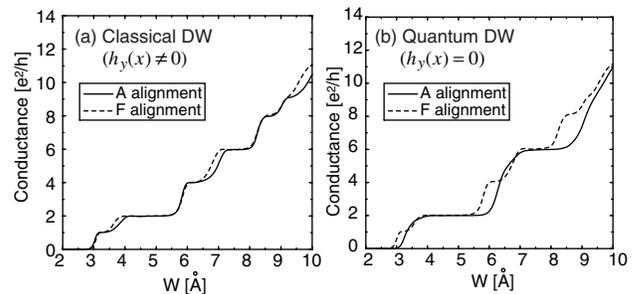}}
      }
  \caption{
    (a) The conductance curve for the A (F) alignment with the classical
    DW is plotted by solid (dotted) linen.  
    (b)The conductance curve for the A (F) alignment with a quantum DW
    is shown. The spin  precession of conduction electron
    in the quantum DW is  forbidden.  In both panels, parameters are
    the same as those in Fig.\protect\ref{fig:PF} (b)}
  \label{fig:AF}
\end{figure}

Let us move on to the effect of the DW for the A alignment.
First, we show how the spin of a conduction electron rotates without
the constriction potential.
We consider a one-dimensional classical DW with $h_{0}=0.5$ eV, and a
spin-up electron incident on the left electrode. 
The reflection probability, 
$|r_{\uparrow\uparrow}|^{2}k_{L}^{\uparrow}/k_{L}^{\uparrow}
+|r_{\downarrow\uparrow}|^{2}k_{L}^{\downarrow}/k_{L}^{\uparrow}$, 
is less than 0.005 even for the $2L=1$ \AA\ and decreases as
the thickness of the DW increases.
The transmission probability  to the spin-down state is as small as
$|t_{\downarrow,\uparrow}|^2(k^{\downarrow}_{R}/k^{\uparrow}_{L}) = 0.16$
for $2L=10$ \AA.
In a PC, however, the geometrical constriction plays a crucial role
in the spin precession through the classical DW.  
In the adiabatic picture, the velocity in the $x$ direction for
channel $n$ is given by $v_{x}^{n}=
\sqrt{(2/m)(E_{F}-E_{n\parallel})}$, where  $E_{n\parallel}$ is the
energy eigenvalue of the transverse mode. 
$E_{n\parallel}$ is a decreasing function of $W$ and  the  channel
$n$ opens when $E_{n\parallel}$ becomes smaller than $E_{F}$.
Electrons transmitting through the channel $n$ have the small
velocity $v_{x}^{n} \ll v_{F}$, where $v_{F}=\sqrt{(2/m)E_{F}}$ is the
Fermi velocity. Electrons track the local exchange
field of the DW adiabatically and feel the constant exchange field
as if they were in the F alignments.  
Therefore, the conductance curve for the A alignment is similar to
that for the F alignment as shown in Fig. \ref{fig:AF}(a).
However, it has been experimentally observed that the sequence of the
quantized conductances is different between F and A alignments:
the conductance plateaus of odd integer multiples of
$e^{2}/h$ appear only for the F alignment \cite{oshima98,ono99}.

This discrepancy can be resolved by considering an atomic scale DW on
the basis of the quantum theory.  
The narrow band $d$-states are
susceptible to disorder due to the small hopping matrix
element and are easily localized \cite{brandbyge_thesis}.
Let us examine the DW in a PC by using the following 
Heisenberg Ferromagnet\cite{schilling77,bader79},
\begin{equation}
{\cal H}_{_{\rm DW}} = - J_{_{\rm DW}}
 \sum_{<i,j>}\vec{S}_{i}\!\cdot\!\vec{S}_{j}
- \alpha\left( \sum_{i\in
  F_{L}} S_{i}^{z}  - \!\!\sum_{i\in F_{R}} S_{i}^{z} \right),
\label{eq:qdw}
\end{equation}
where $\vec{S}_{i}$ is the localized $3d$ spin at site $i$.  We
consider the nearest-neighbor interactions with coupling constant
$J_{_{\rm DW}}$ and neglect the anisotropy.  The second term contains
the interactions of spins in the left (right) electrode $F_{L(R)}$
with the coupling constant $\alpha$.  We assume that $S_{i}=1$ for Ni
PC.  The eigenstates of the quantum DW given by Eq. (\ref{eq:qdw}) are
labeled by the $z$ component of the total spin
$S^{z}\equiv\sum_{i}S_{i}^{z}$ and the ground state is $S^{z}=0$.  The
exchange energy of the effective one-electron Hamiltonian
$-\vec{h}(x)\cdot\vec{\sigma}$ is expressed as
$-J\sum_{i}\vec{s}\cdot\langle\vec{S}_{i}\rangle$, where $\vec{s}$
is the spin of a conduction electron and 
$\langle\vec{S}_{i}\rangle$ represents the expectation value of the
localized spin $\vec{S}_{i}$ and $J$ is the corresponding coupling
constant\cite{berger84}.
Since the DW consists of a few atoms
and is strongly pinned by the geometrical constriction, the DW has a
large excitation energy and the expectation value
$\langle\vec{S}_{i}\rangle$ is evaluated in the ground state with $S^{z}=0$.
We studied the $N \le 8$ site one-dimensional DW with $S=1$ by
numerically diagonalizing the Hamiltonian $\cal{H}_{_{\rm DW}}$.  We
find that the $\langle S_{i}^{z} \rangle$ is well fitted by $h_{z}(x)$
in Eq. (\ref{eq:hz}) and the
excitation energy for $S^{z}=\pm 1$ is large.  For example, the excitation
energy is 1.1 $J_{_{\rm DW}}$ for $N=4$ and $J_{_{\rm DW}}=\alpha$.  One
crucial property of such an atomic scale quantum DW is that the
expectation values $\langle S_{i}^{x} \rangle= \langle S_{i}^{y}
\rangle =0$ for all sites $i$, because the
operators $S_{i}^{x}$ and $S_{i}^{y}$ have only nonzero matrix
elements between states of different eigenvalue of
$S_{z}$\cite{schilling77}.  Therefore, the spin-mixing term vanishes,
{\it i. e.}, $-h_{y}(x)\sigma_{y}=-J\sum_{i}\vec{s}\cdot\langle
S_{i}^{y}\rangle=0$ in Eq. (\ref{eq:hamiltonian1}) for the atomic
scale quantum DW.  The spin of the conduction
electrons cannot rotate in the DW.

The conductance curve for the A alignment with a quantum DW are plotted 
in Fig. \ref{fig:AF} (b). 
The sequence of the quantized conductances is clearly different
between F and A alignments.
In the adiabatic picture, the number of transmitting channels is the
same for spin-up and spin-down electrons because the exchange energy
-$h_{z}(x)\sigma_{z}$ is an odd function of $x$.  The
sequence of the quantized conductances is the same as that for a
nonmagnetic PC shown in Fig. \ref{fig:PF} (a).  The
conductance curve shows clear plateaus at $G=2e^{2}/h$ and 6 $e^{2}/h$
and the plateaus at odd
integer multiples of $e^{2}/h$ disappear.
Comparing Figs. \ref{fig:AF} (a) and \ref{fig:AF} (b), we conclude
that the recent experimental results that the sequence of the quantized
conductances is different between A and F alignments
\cite{oshima98,ono99} is the direct consequence
of the fact that the spin of conduction electrons cannot rotate
in the atomic scale quantum DW.

\begin{figure}
  \epsfxsize=0.95\columnwidth \centerline{\hbox{ \epsffile{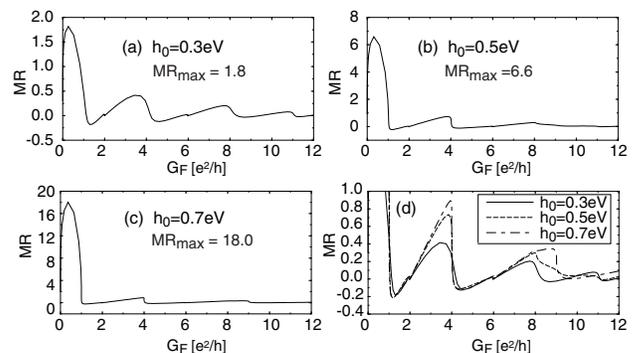}}}
  \caption{The Magnetoresistances(MR) for the PC with $h_{0}=0.3$,
    0.5 and 0.7 eV are shown in panels (a), (b) and (c), respectively.
    The holizontal axis represents the conductance for the F alignment.
    The enlarged views of MR are shown in the panel (d).}
  \label{fig:mr}
\end{figure}

In Figs. \ref{fig:mr}(a)-(d), we show the magnetoresistance(MR)
calculated by the formula: MR$=(G_{F}/G_{A})-1$, where $G_{F(A)}$ is
the total conductance for the F (A) alignment.
We find the strong enhancement of the MR at $G_{F}=1$($W\simeq
3$\AA)\, where the first transmitting channel opens for the F alignment
but channels for the A alignment are hardly transmittive.
The point is that the DW makes the number of
transmitting channels different between F and A alignments as shown in
Fig. \ref{fig:AF} (b).  Note that the conductance plateau at 
integer multiples of $e^{2}/h$ means that the scattering intrinsic
to the DW \cite{cabrera74,hoof99} is negligible.
The magnetoresistance,
i.e., the difference of conductances between F and A alignments
increases as the exchange field $h_{0}$ increases.  
The maximum values of MR,
${\rm MR_{max}}$, for $h_{0}=$ 0.3, 0.5 and 0.7 eV are respectively 1.8,
6.6 and 18.0.
Our results explain the MR enhancement 
observed by Garc{\`\i}a {\it et al.}\cite{garcia99}.
The enhancement of MR is expected to be large if 
magnetic-metals with large exchange field such as Co and Fe are used.
We also find that the MR oscillates with $G_{F}$ as shown
in Fig. \ref{fig:mr} (d).
Since the difference of the conductances or the
number of transmitting channels between the A and F alignments does
not increase with $G_{F}$ as shown in Fig. \ref{fig:AF} (b),  the
magnitude of the oscillation decreases with $G_{F}$.  

In conclusion, we have studied the electron transport through a
magnetic PC with  special attention to the effect of an atomic scale DW.  
We show the sequence of the quantized conductances
depends on the relative orientation of magnetizations between left and
right electrodes.  The quantized conductance of odd integer multiples
of $e^{2}/h$ appears only for the F alignment.  For the A alignment, the
unit of the conductance quantization is even integer multiples of $e^{2}/h$
since the spin precession of the conduction electron is forbidden in
the atomic scale DW.  We also show that the magnetoresistance is strongly
enhanced in the narrow PC and oscillates with the width of the
constriction.  
The realistic band structures may affect the reflection probability
intrinsic to the DW and thus the MR\cite{hoof99}.
In magnetic PC, however, such effect is negligible and 
what is of most importance in understanding electron transport in a magnetic
PC is the fact that the DW makes the number of
transmitting channels different between F and A alignments.

We acknowledge T. Ono, S. Mitani, K. Miyano, M. Kohno and M. Tsukada
for valuable discussions.  This work is supported by a Grant-in-Aid for
Scientific Research Priority Area for Ministry of Education, Science
and Culture of Japan, a Grant for the Japan Society for Promotion of
Science and NEDO Japan.



\end{document}